\documentclass[aps,preprint]{revtex4}
\usepackage{amssymb,epsf}
\usepackage{amsmath}
\usepackage{graphicx}

\begin{document}
\title{Asymptotically Flat Radiating Solutions in Third Order Lovelock Gravity}
\author{M. H. Dehghani$^{1,2}$\footnote{email address:
mhd@shirazu.ac.ir}, N. Farhangkhah$^{1}$}

\affiliation{$^1$Physics Department and Biruni Observatory,
College of Sciences, Shiraz
University, Shiraz 71454, Iran\\
$^2$Research Institute for Astrophysics and Astronomy of Maragha
(RIAAM), Maragha, Iran}
\begin{abstract}
In this paper, we present an exact spherically symmetric solution of third order
Lovelock gravity in $n$ dimensions which describes
the gravitational collapse of a null dust fluid. This solution is asymptotically
(anti-)de Sitter or flat depending on the choice of the cosmological constant.
Using the asymptotically flat solution for $n \geq 7$ with a power-law
form of the mass as a function of the null coordinate, we present a model for a gravitational
collapse in which a null dust fluid radially injects into an initially flat and empty region. It is found
that a naked singularity is inevitably formed whose strength is different for the $n = 7$
and $n \geq 8$ cases. In the $n=7$ case, the limiting focusing condition for
the strength of curvature singularity is satisfied.
But for $n \geq 8$, the strength of curvature singularity depends on the rate of increase of mass
of the spacetime. These considerations show that
the third order Lovelock term weakens the strength of the curvature singularity.
\end{abstract}

\pacs{04.20.Dw, 04.50.+h, 04.20.Jb, 04.70.Bw}
\maketitle
\section{Introduction}

The problem of gravitational collapse in general relativity is one of the
unsolved problems in gravitation physics. One would like to know whether,
and under what initial conditions, gravitational collapse results in black
hole formation. In particular, one would like to know if there are physical
collapse solutions that lead to naked singularities. In 1939, Oppenheimer and
Snyder \cite{Oppen} studied, as an idealized model of gravitational
collapse, the solution which corresponds to a homogeneous spherically
symmetric dust cloud. By analyzing the behavior of the outgoing light rays
they were led to black hole idea. In general relativity, singularity
theorems were proved stating that spacetime singularities inevitably appear
under general situations and physical energy conditions \cite{Haw}. However,
the nature of the singularities remains an outstanding unresolved question.
In this context, a cosmic censorship hypothesis (CCH) was proposed by
Penrose, which states that curvature singularities in asymptotically flat
spacetimes are always shrouded by event horizons; in other words, there are
no naked singularities formed in physical gravitational collapse \cite
{conjecture}. The weak CCH allows for the occurrence of locally naked
singularities but not globally naked ones, whereas the strong CCH does not
allow either. However, there are a number of important results which assume
the truth of this hypothesis, such as the area theorem of black holes, and soon after
it some counterexamples to this hypothesis were found. One of those is the
generic occurrence of naked singularities in the null dust collapse shown by
Vaidya. In 1959, Vaidya found a solution that represents an imploding
(exploding) null dust fluid with spherical symmetry \cite{Vaid}. Since then
this solution has been studied in gravitational collapses by many authors.
In particular, it has been showed that this solution can give rise to the
formation of the naked singularity, and after that some other counterexamples
to the cosmic censorship hypothesis were provided \cite{counter}.

There exist many possible generalizations of the Vaidya-type metric. The (anti-)de Sitter [(A)dS]
Vaidya-type solutions in Einstein gravity have been worked
widely until now in \cite{AdS} and the references therein. In these papers,
the asymptotically (A)dS spherically symmetric Vaidya-type solutions of
Einstein gravity with cosmological constant have been found, and the
existence of naked singularity has been investigated. These papers show
that whether the spacetime is asymptotically flat or not has no any effect
on the occurrence of a locally naked singularity. Also the effects of the
dimensionality of spacetime on the weak CCH has been investigated \cite
{VaidND}.

In recent years a renewed interest has grown in higher-order
gravity, which involves higher derivative curvature terms. Among
them, the second order Lovelock gravity or the so-called
Einstein-Gauss-Bonnet gravity and the third order Lovelock gravity
are of particular interest because of their special features.
Exact solutions of the former can be found in \cite {deser,GB} and
the latter in \cite{shah,mann}. Also, the possibility of a dark
energy universe emerging from an action with higher-order
curvature terms has been investigated \cite{Od}. Here we want to
investigate the effects of higher-order Lovelock terms on CCH for
those solutions which have a general relativistic limit. The effects
of second order Lovelock terms on the Vaidya-type solutions have
been investigated in \cite{kobaya,Maed1}. These papers show that
the appearance of a second order Lovelock term in the field
equations has no effect on the occurrence of locally naked
singularity, while it has some effects on the strength of the
curvature. In this paper, our aim is to consider the effect of the
third order Lovelock term on the gravitational collapse of the
null fluid, the occurrence of a naked singularity, and the strength
of the curvature. The Vaidya-type solution of dimensionally
continued Lovelock gravity has been found in \cite{maeda}. This
solution is asymptotically AdS and contains only one fundamental
constant. Indeed, this solution cannot show the effect of
higher-order Lovelock terms explicitly on the weak CCH, which is
proposed for asymptotically flat spacetimes of Einstein gravity.
Here we find asymptotically flat and (A)dS Vaidya-type solutions
of third order Lovelock gravity with four fundamental constants
(the cosmological constant and the three Lovelock coefficients).

The outline of our paper is as follows. In Sec. \ref{Vai}, we introduce the
Vaidya-type solutions of third order Lovelock gravity and obtain the $n$%
-dimensional solution of a spacetime outside a radiating star with the
energy-momentum tensor of a null fluid. In Sec. \ref{Exist}, we show that the final fate
of this spacetime is a naked singularity. Section \ref{Stren} is devoted to the investigation
of the strength of the curvature singularity for different rates of increase of
the mass of the spacetime in an arbitrary dimension. We finish our paper with
some concluding remarks.

\section{Vaidya-type Solution in third order Lovelock gravity \label{Vai}}
A natural generalization of general relativity in higher-dimensional
spacetimes with the assumption of Einstein, that the left hand side of the
field equations is the most general symmetric conserved tensor containing no
more than second derivatives of the metric, is the Lovelock theory. The
gravitational field equations of third order Lovelock gravity may be written
as
\begin{equation}
\Lambda' g_{\mu \nu }+G_{\mu \nu }^{(1)}+\alpha _{2}^{\prime
}G_{\mu \nu }^{(2)}+\alpha _{3}^{\prime }G_{\mu \nu
}^{(3)}=\kappa^2 _{n}T_{\mu \nu }, \label{Geq}
\end{equation}
where $\Lambda' $ is the cosmological constant, $\alpha
_{i}^{\prime }$'s are Lovelock coefficients which are assumed to
be positive throughout the paper, $T_{\mu \nu }$ is the
energy-momentum tensor of the matter, $G_{\mu \nu }^{(1)}$ is the
Einstein tensor, and $G_{\mu \nu }^{(2)}$ and $G_{\mu \nu }^{(3)}$
are the second and third order Lovelock tensors, respectively, given as
\begin{equation}
G_{\mu \nu }^{(2)}=2(R_{\mu \sigma \kappa \tau }R_{\nu }^{\phantom{\nu}%
\sigma \kappa \tau }-2R_{\mu \rho \nu \sigma }R^{\rho \sigma }-2R_{\mu
\sigma }R_{\phantom{\sigma}\nu }^{\sigma }+RR_{\mu \nu })-\frac{1}{2}%
\mathcal{L}_{2}g_{\mu \nu },  \label{Love2}
\end{equation}
\begin{eqnarray}
G_{\mu \nu }^{(3)} &=&-3(4R^{\tau \rho \sigma \kappa }R_{\sigma \kappa
\lambda \rho }R_{\phantom{\lambda }{\nu \tau \mu}}^{\lambda }-8R_{%
\phantom{\tau \rho}{\lambda \sigma}}^{\tau \rho }R_{\phantom{\sigma
\kappa}{\tau \mu}}^{\sigma \kappa }R_{\phantom{\lambda }{\nu \rho \kappa}%
}^{\lambda }+2R_{\nu }^{\phantom{\nu}{\tau \sigma \kappa}}R_{\sigma \kappa
\lambda \rho }R_{\phantom{\lambda \rho}{\tau \mu}}^{\lambda \rho }  \notag \\
&&-R^{\tau \rho \sigma \kappa }R_{\sigma \kappa \tau \rho }R_{\nu \mu }+8R_{%
\phantom{\tau}{\nu \sigma \rho}}^{\tau }R_{\phantom{\sigma \kappa}{\tau \mu}%
}^{\sigma \kappa }R_{\phantom{\rho}\kappa }^{\rho }+8R_{\phantom
{\sigma}{\nu \tau \kappa}}^{\sigma }R_{\phantom {\tau \rho}{\sigma \mu}%
}^{\tau \rho }R_{\phantom{\kappa}{\rho}}^{\kappa }  \notag \\
&&+4R_{\nu }^{\phantom{\nu}{\tau \sigma \kappa}}R_{\sigma \kappa \mu \rho
}R_{\phantom{\rho}{\tau}}^{\rho }-4R_{\nu }^{\phantom{\nu}{\tau \sigma
\kappa }}R_{\sigma \kappa \tau \rho }R_{\phantom{\rho}{\mu}}^{\rho
}+4R^{\tau \rho \sigma \kappa }R_{\sigma \kappa \tau \mu }R_{\nu \rho
}+2RR_{\nu }^{\phantom{\nu}{\kappa \tau \rho}}R_{\tau \rho \kappa \mu }
\notag \\
&&+8R_{\phantom{\tau}{\nu \mu \rho }}^{\tau }R_{\phantom{\rho}{\sigma}%
}^{\rho }R_{\phantom{\sigma}{\tau}}^{\sigma }-8R_{\phantom{\sigma}{\nu \tau
\rho }}^{\sigma }R_{\phantom{\tau}{\sigma}}^{\tau }R_{\mu }^{\rho }-8R_{%
\phantom{\tau }{\sigma \mu}}^{\tau \rho }R_{\phantom{\sigma}{\tau }}^{\sigma
}R_{\nu \rho }-4RR_{\phantom{\tau}{\nu \mu \rho }}^{\tau }R_{\phantom{\rho}%
\tau }^{\rho }  \notag \\
&&+4R^{\tau \rho }R_{\rho \tau }R_{\nu \mu }-8R_{\phantom{\tau}{\nu}}^{\tau
}R_{\tau \rho }R_{\phantom{\rho}{\mu}}^{\rho }+4RR_{\nu \rho }R_{%
\phantom{\rho}{\mu }}^{\rho }-R^{2}R_{\nu \mu })-\frac{1}{2}\mathcal{L}%
_{3}g_{\mu \nu }.  \label{Love3}
\end{eqnarray}
In Eqs. (\ref{Love2}) and (\ref{Love3}) $\mathcal{L}_{2}=R_{\mu \nu \gamma
\delta }R^{\mu \nu \gamma \delta }-4R_{\mu \nu }R^{\mu \nu }+R^{2}$ is the
Gauss-Bonnet Lagrangian and
\begin{eqnarray}
\mathcal{L}_{3} &=&2R^{\mu \nu \sigma \kappa }R_{\sigma \kappa \rho \tau }R_{%
\phantom{\rho \tau }{\mu \nu }}^{\rho \tau }+8R_{\phantom{\mu \nu}{\sigma
\rho}}^{\mu \nu }R_{\phantom {\sigma \kappa} {\nu \tau}}^{\sigma \kappa }R_{%
\phantom{\rho \tau}{ \mu \kappa}}^{\rho \tau }+24R^{\mu \nu \sigma \kappa
}R_{\sigma \kappa \nu \rho }R_{\phantom{\rho}{\mu}}^{\rho }  \notag \\
&&+3RR^{\mu \nu \sigma \kappa }R_{\sigma \kappa \mu \nu }+24R^{\mu \nu
\sigma \kappa }R_{\sigma \mu }R_{\kappa \nu }+16R^{\mu \nu }R_{\nu \sigma
}R_{\phantom{\sigma}{\mu}}^{\sigma }-12RR^{\mu \nu }R_{\mu \nu }+R^{3}
\end{eqnarray}
is the third order Lovelock Lagrangian. In order to have the contribution of
all of the above terms in the field equation, the dimension of the spacetime
should be equal or larger than $7$.

Here we are looking for Vaidya-type solutions of third order Lovelock
gravity. Thus, the only nonvanishing component of the energy-momentum tensor
is $T_{v}\,^{r}$. The metric of the $n$-dimensional spherically symmetric
Vaidya-type solution may be written as

\begin{equation}
ds^{2}=-f(r,v)dv^{2}+2\epsilon drdv+r^{2}d\Omega _{n-2}^{2},  \label{metric1}
\end{equation}
where $v\in (-\infty ,+\infty )$ is a null coordinate which represents
advanced Eddington time and is ingoing (outgoing) for $\epsilon =+1(-1)$, $%
0\leq r<\infty $ is the radial coordinate, and $d\Omega _{n-2}^{2}$\ is the
line element of the $(n-2)$-dimensional unit sphere:
\begin{equation*}
d\Omega _{n-2}^{2}=d\theta
_{1}^{2}+\sum\limits_{i=2}^{n-2}\prod\limits_{j=1}^{i-1}\sin ^{2}\theta
_{j}d\theta _{i}^{2}.
\end{equation*}
We want to obtain the Vaidya-type solutions of third order Lovelock gravity.
In this case, the energy-momentum tensor for the directional flow of
radiation in empty space is
\begin{mathletters}
\begin{equation}
T_{a}\,^{b}=\sigma (r,v)k_{a}k^{b},  \label{Tab}
\end{equation}
where $k_{a}=-\partial _{a}v$ and $\sigma (r,v)$ is the density of flowing
radiation. The only nonvanishing component of $T_{a}\,^{b}$ is $T_{v}\,^{r}$, and
therefore the field equations (\ref{Geq}) reduce to the
following differential equations for $f(r,v)$:
\end{mathletters}
\begin{eqnarray}
&&(n-1)\Lambda r^{6}+\left\{ \alpha _{3}r[f(r,v)-1]^{2}-2\alpha
_{2}r^{3}[f(r,v)-1]+r^{5}\right\} f^{\prime }+
\notag \\
&&\frac{n-7}{3}\alpha _{3}[f(r,v)-1]^{3}-(n-5)\alpha
_{2}r^{2}[f(r,v)-1]^{2}+(n-3)r^{4}[f(r,v)-1]=0,  \label{E11}\\
&& -\frac{(n-2)}{2r^{5}}\left\{ \alpha _{3}\left[ f(r,v)-1%
\right] ^{2}-2\alpha _{2}r^{2}\left[ f(r,v)-1\right]
+r^{4}\right\} \dot{f} =\kappa^2 _{n}T_{v}\,^{r},  \label{E12}
\end{eqnarray}
where the prime and the dot denote the derivatives with respect to the coordinates $%
r $ and $v$, respectively and we define $\Lambda^{\prime}
=2\Lambda/(n-1)(n-2)$, $\alpha _{2}^{\prime }=\alpha
_{2}/(n-3)(n-4)$, and $\alpha _{3}^{\prime }=\alpha
_{3}/3(n-3)...(n-6)$ for simplicity. Equation (\ref{E11}) has one
real and two complex solutions. The only real solution of Eq.
(\ref{E11}) which reduces to the solution of Einstein gravity in
the limit of $\alpha_2=\alpha_3=0$ can be written as

\begin{eqnarray}
f(r,v) &=&1+\frac{\alpha _{2}r^{2}}{\alpha _{3}}\left\{
1-C^{1/3}(r,v)+\gamma ^{1/3}C^{-1/3}(r,v)\right\} ,  \notag \\
C(r,v) &=&\left( \sqrt{\gamma +k^{2}(r,v)}+k(r,v)\right) ,  \notag \\
k(r,v) &=&-1+\frac{3\alpha _{3}}{2\alpha _{2}^{2}}+\frac{3\Lambda
\alpha _{3}^{2}}{2\alpha _{2}^{3}}+\frac{3 \alpha _{3}^{2}}{%
2\alpha _{2}^{3}}\frac{M(v)}{r^{n-1}},\text{ \ \ \ \ \ }\gamma =\left( \frac{%
\alpha _{3}-\alpha _{2}^{2}}{\alpha _{2}^{2}}\right) ^{3},  \label{fstat}
\end{eqnarray}
where $M(v)$ is the measure of the mass of the spacetime (see Appendix A).
One should note that $\gamma +k^2(r,v)$ is larger
than $\gamma +k^2(\infty ,v)$, and the latter should be positive
or at least zero due to the reality of the metric function
$f(v,r)$ everywhere. This concludes that $\alpha_3> 3\alpha_2^2/4$.
The metric given in Eqs. (\ref{metric1}) and (\ref{fstat}) is
asymptotically flat for $\Lambda =0$ and AdS and dS for negative
and positive values of $\Lambda $, respectively. In this paper, we
are interested in asymptotically flat Vaidya-type solutions with
a general relativistic limit in order to consider CCH in
higher-order Lovelock
gravity. This solution reduces to an $n$-dimensional Vaidya-type solution as $%
\alpha _{2}$ and $\alpha _{3}$ vanish \cite{Vaid}. The case that $M=$const.
reduces to the static solution of third order Lovelock gravity, which is
considered in \cite{shah}.

The only nonvanishing component of stress energy tensor can be found from Eq. 
(\ref{E12}) as
\begin{equation}
T_{v}\,^{r}=\frac{(n-2)}{2\kappa^2 _{n}r^{n-2}} \dot{M}. \label{EMten}
\end{equation}
It is worthwhile to note that the dependence of the energy-momentum tensor given
in Eq. (\ref{EMten}) on $v$ and $r$ is the same as that of
Vaidya-type solutions of Einstein \cite{VaidND}, Gauss-Bonnet \cite{kobaya}
or dimensionally continued Lovelock \cite{maeda} gravities. 

\section{The Existence and Nature of A Naked Singularity \label{Exist}}
In this and the next sections, we study the gravitational collapse of a null
dust fluid in third order Lovelock gravity and compare it with that in
Einstein relativity \cite{VaidND}. In order to investigate CCH in third
order Lovelock gravity, we consider asymptotically flat solutions ($\Lambda
=0$). The physical situation here is that of a radial influx of null fluid
in an initially empty region of the higher-dimensional Minkowskian
spacetime. The first shell arrives at $r=0$ at time $v=0$ and the final shell at $%
v=v_{f}$. A central singularity of growing mass is developed at $r=0$. For $%
v<0$, we have $m(v)=0$, i.e., higher-dimensional Minkowskian spacetime, and
for $v>v_{f}$, $M(v)=M_{f}$ is a positive constant, and we have the static
solution of third order Lovelock gravity considered in \cite{shah}. We
choose a power-law growth of mass as
\begin{equation*}
M(v)=\left\{
\begin{array}{ll}
0 & v<0 \\
M_{0}v^{p}\text{ } & 0\leq v \leq v_{f} \\
M=M_{0}v_{f}^{p} & v>v_{f}
\end{array}
\right.
\end{equation*}
In order to consider the existence of a physical singularity, we compute the
Kretschmann scalar. It is a matter of calculations to show that the
Kretschmann scalar diverges for $r\rightarrow 0$ as
\begin{equation*}
K=O\left( \frac{v^{2p/3}}{r^{2(n-1)/3}}\right) ,
\end{equation*}
which shows that there is a singularity at $r=0$ for $v\geq 0$.

The nature of the singularity (to be naked or hidden) can be
characterized by the existence of radial null geodesics coming out
of the singularity. It can be shown that if a future-directed
radial null geodesic does not emanate from the singularity, a
future-directed causal (excluding radial null) geodesic does not,
too \cite{maeda}. So we consider here only the future-directed
outgoing radial null geodesics which satisfies
\begin{equation}
\frac{dr}{dv}=\frac{f}{2}.  \label{diff}
\end{equation}
The region with $f<0$ is the trapped region, and a hypersurface $f=0$
represents the trapping horizon \cite{Hay}. The radius of the apparent
horizon as a function of $v$ is given by the following equation:
\begin{equation}
M(v)=M_{0}v^{p}=\frac{\alpha _{3}}{3}r^{n-7}+\alpha _{2}r^{n-5}+r^{n-3}.
\label{mass}
\end{equation}
Along the trapping horizon we have:

\begin{equation}
ds^{2}=\frac{6pM_{0}v^{p-1}}{(n-7)\alpha _{3}r^{n-8}+3(n-5)\alpha
_{2}r^{n-6}+3(n-3)r^{n-4}}dv^{2},
\end{equation}
and hence it is spacelike for $v>0$ and $r>0$ for positive Lovelock coefficients. It is a future outer
trapping horizon, which is a local definition of black hole (see Appendix B). It
is seen from (\ref{mass}) that only the point $r=v=0$ may be a naked
singularity for $n\geq 8$, while the central singularity $(r=0$, $v>0)$ is
naked for $0\leq v\leq v_{ah}$ in the case of $n=7$, where $v_{ah}$ is given
as
\begin{equation}
3M_{0}v_{ah}^{p}=\alpha _{3}.
\end{equation}

In order to show the existence of a null geodesic emanating from the
singularity, we adopt the fixed-point method \cite{christo}. We introduce a
new coordinate$\mathcal{\ \vartheta }=r/(v-v_{0})$, where $v_{0}$ is a
constant which is in the range $0\leq v_{0}\leq v_{ah}$ when $\ n=7,$ while
it is zero for $n\geq 8$. Then the null geodesic equation (\ref{diff})
becomes
\begin{equation}
\frac{dr}{dv}=\frac{d\mathcal{\vartheta }}{dv}(v-v_{0})+\mathcal{\vartheta =}%
\frac{1}{2}\left\{ 1+\frac{\alpha _{2}r^{2}}{\alpha _{3}}\left[
1-C(v,r)^{1/3}+\gamma ^{1/3}C(v,r)^{-1/3}\right] \right\} ,
\end{equation}
which may be written as
\begin{equation}
\frac{d\mathcal{\vartheta }}{dv}+\frac{1}{(v-v_{0})}(\mathcal{\vartheta }%
-\eta )=\eta \Psi ,  \label{eq3}
\end{equation}
where we have introduced the parameter $\eta $ in the range $0<\eta <\infty $
and $\Psi $ is
\begin{equation*}
\Psi =-\frac{1}{(v-v_{0})}+\frac{1}{2\eta (v-v_{0})}\left\{ 1+\frac{\alpha
_{2}(\mathcal{\vartheta }(v-v_{0}))^{2}}{\alpha _{3}}\left[
1-C(v,r)^{1/3}+\gamma ^{1/3}C(v,r)^{-1/3}\right] \right\}.
\end{equation*}
If $\eta $\ is chosen to be\bigskip
\begin{equation*}
\eta _{0}=\frac{1}{2}\left\{ 1-\left( \frac{3M_{0}v_{0}^{p}}{\alpha _{3}}%
\right) ^{1/3}\right\}
\end{equation*}
for \ $n\geq 7$, $\Psi $ is at least $C^{1}$\ for $v\geq v_{0}$ and $%
\mathcal{\vartheta }>0$. In this case, Eq. (\ref{eq3}) has a nonnegative
unique solution which satisfies $\mathcal{\vartheta }\mathcal{(}v_{0}%
\mathcal{)=}\eta _{0}$ and is continuous at $v=v_{0}$ \cite{christo}. This
solution represents an outgoing light ray which emanates from the singular
point at the center and which either intersects the apparent horizon or
goes to future null infinity \cite{christo}. Thus, the solution may be
interpreted as a naked singularity.

\section{Strength of the Singularity \label{Stren}}
The strength of a singularity, which is the measure of its destructive
capacity, is the most important feature. Following the paper \cite{Clarke} we
consider the null geodesics affinely parametrized by $\lambda $ and
terminating at a shell focusing singularity $r=v=\lambda =0$. Then the solution satisfies
a strong curvature condition (SCC) \cite{tip} if
\begin{equation}
\lim_{\lambda \rightarrow 0}\lambda ^{2}\Phi >0,  \label{CCF}
\end{equation}
and limiting focusing condition (LFC) \cite{Krol} if
\begin{equation}
\lim_{\lambda \rightarrow 0}\lambda \Phi >0,  \label{LFC}
\end{equation}
where
\begin{equation*}
\Phi \equiv R_{\mu \nu }k^{\mu }k^{\nu }.
\end{equation*}
Using the fact that $dr/d\lambda =(dv/d\lambda )f/2$, one can show that
\begin{equation}
\Phi =-\frac{2(n-2) \dot{f}}{rf^2}\left( \frac{dr}{d\lambda }%
\right) ^{2},  \label{psi}
\end{equation}
and the radial null geodesic satisfies the differential equation
\begin{equation}
\frac{d^{2}r}{d\lambda ^{2}}\simeq \frac{2 \dot{f}}{f^{2}}\left(
\frac{dr}{d\lambda }\right) ^{2}.  \label{d2r}
\end{equation}

\subsection{$p=n-3$ case:}
In this subsection, in order to compare our results with the results of
the curvature strength of $n$-dimensional solutions of Einstein and Gauss-Bonnet
gravities investigated in \cite{VaidND} and \cite{kobaya}, respectively, we
consider the case $p=n-3$. First, we consider the seven-dimensional solutions
of third order Lovelock gravity. In this case
\begin{eqnarray}
&& \underset{r\rightarrow 0}{\lim }f(r,v) = 2\eta _{0},  \label{limfv} \\
&& \underset{r\rightarrow 0}{\lim } \dot{f} =-\frac{4}{3}\left(
\frac{3M_{0}v_{0}}{\alpha _{3}}\right) ^{1/3},  \label{limf}
\end{eqnarray}
and therefore the radial null geodesic equation near $r=v=0$ can be written
as

\begin{equation*}
\frac{d^{2}r}{d\lambda ^{2}}+\frac{2}{3\eta _{0}^{2}}\left( \frac{%
3M_{0}v_{0}}{\alpha _{3}}\right) ^{1/3}\left( \frac{dr}{d\lambda }%
\right) ^{2}\simeq 0,
\end{equation*}
with the solution
\begin{equation}
r\simeq \frac{3\eta _{0}^{2}}{2}\left( \frac{\alpha _{3}}{3M_{0}v_{0}}%
\right) ^{1/3}\ln (\lambda +1).  \label{rl}
\end{equation}
Using Eqs. (\ref{psi}), (\ref{limfv}), (\ref{limf}) and (\ref{rl}), one
finds that the limit of $\lambda \Phi $ is positive while the limit of $%
\lambda ^{2}\Phi $ is zero as $\lambda $ goes to zero, and therefore only
LFC is satisfied along a radial null geodesic. That is, along the radial
null geodesics the strong curvature condition is not satisfied. Thus, the
singularity of seven-dimensional third order Lovelock gravity is weaker than
that of Vaidya-type solutions of Einstein gravity \cite{VaidND}, while it has
the strength as in the case of Gauss-Bonnet gravity \cite{kobaya}.

Second, we consider $n\geq 8$ for which the limit of $f(r,v)$ is:
\begin{equation}
\underset{r\rightarrow 0}{\lim }f(r,v) =1,  \label{limfv8}
\end{equation}
Using Eq. (\ref{diff}), one obtains $v \simeq 2r$ for small $r$ and therefore
the limit of $\dot{f}$ becomes
\begin{equation}
\underset{r\rightarrow 0}{\lim } \dot{f} =-\left(
\frac{2^{n}M_{0}r}{9\alpha _{3}}\right) ^{1/3}=0.  \label{limf8}
\end{equation}
In this case, one can show that the limits of both $\lambda \Phi $ and $%
\lambda ^{2}\Phi $ are zero as $\lambda $ $\rightarrow 0$. In other words,
neither the SCC nor the LFC is satisfied, and therefore the third order Lovelock
term weakens the strength of the singularity.

\subsection{$p=n-4$ case:}
Again, it is easy to show that along the radial
null geodesics only the LFC is satisfied for a seven-dimensional solution.
For $n \geq 8$, the radial null geodesic equation near $r=v=0$ can
be written as
\begin{equation*}
\frac{d^{2}r}{d\lambda ^{2}}+\frac{2p}{3}\left( \frac{3M_{0}}{\alpha _{3}}%
\right) ^{1/3}\left( \frac{dr}{d\lambda }\right) ^{2}\simeq 0,
\end{equation*}
with the solution
\begin{equation}
r\simeq \frac{3}{2p}\left( \frac{\alpha _{3}}{3M_{0}}\right) ^{1/3}\ln
(\lambda +1).  \label{r2}
\end{equation}
Using Eqs. (\ref{psi}), (\ref{limfv8}), (\ref{limf8}) and (\ref{r2}), one
finds that the limit of $\lambda \Phi $ is positive while the limit of $%
\lambda ^{2}\Phi $ is zero as $\lambda $ goes to zero, and therefore only
the LFC is satisfied along a radial null geodesic.

\section{Closing Remarks}

Considering the spherically symmetric gravitational collapse of a null dust
fluid in $n\geq 7$ dimensions, we have obtained an exact solution in third
order Lovelock gravity with four fundamental constants which is
asymptotically (A)dS or flat depending on the choice of the cosmological
constant. This solution reduces to the Vaidya-type solution of Gauss-Bonnet
gravity \cite{kobaya} as the third order Lovelock term is turned off, and
reduces to the $n$-dimensional Vaidya-type solution of Einstein gravity for $%
\alpha _{1}=\alpha _{2}=0$ \cite{VaidND}. We applied the solution
to the situation in which a null dust fluid radially injects into
an initially flat region with the rate $M(v)=M_{0}v^{p}$, and
investigated the effects of the third order Lovelock term on the
final fate of the gravitational collapse. We found that, as in the
case of Gauss-Bonnet gravity, a naked singularity is inevitably
formed. In the general relativistic case, a naked singularity will
form only when $M_{0}$ takes a sufficiently small value, and
therefore turning on the Lovelock terms worsens the situation from
the viewpoint of CCH. Furthermore, as in the case of Gauss-Bonnet
gravity, the third order Lovelock term changes the nature of the
singularity and the whole picture of gravitational collapse
drastically. The picture of the gravitational collapse for $n = 7$
is quite different from that for $n \geq 8$, as well as the
general relativistic case for $n \geq 4$ and the Gauss-Bonnet case
for $n \geq 6$, a massless ingoing null naked singularity is
formed. On the other hand, for the special case $n = 7$, as in the
case of five-dimensional Gauss-Bonnet gravity, a massive timelike
naked singularity is formed. As can be seen from the result of
Ref. \cite{Maed1} and this paper, the nature of the naked
singularity is quite different for $n=2i_{\texttt{max}}+1$ and
$n>2i_{\texttt{max}}+1$, where $i_{\texttt{max}}+1$ is the largest
order of Lovelock term. Thus, we conjecture that \emph{the naked
singularity is massive in} $n=2i_{\texttt{max}}+1$, \emph{while it
is massless in higher dimensions}.

Although naked singularities are inevitably formed in third order Lovelock
gravity, the strength of the singularity is different in the cases of Einstein,
Gauss-Bonnet, and third order Lovelock gravities. In seven dimensions, the LFC
is satisfied for the solutions of Gauss-Bonnet and third order Lovelock
gravities, while the SCC is satisfied in Einstein gravity. Thus, the strength of
singularities of seven-dimensional solutions of Gauss-Bonnet and third order
Lovelock gravities is weaker than that of Einstein gravity.  In $n\geq 8$ dimensions
for $p=n-3$, the solution satisfies neither the LFC nor the SCC in third order Lovelock gravity,
while the LFC is satisfied in Gauss-Bonnet gravity for the $p=n-3$. Thus,
in $n\geq 8$ dimensions, turning on the third order Lovelock gravity weakens
the strength of the naked singularity for the $p=n-3$ case. We also considered
the strength of singularity in third order Lovelock gravity for $p=n-4$, and
found that the LFC is satisfied along a radial null geodesic in $n\geq 8$
dimensions. Thus, we conjecture that
\emph{the higher-order Lovelock terms weaken the strength of the singularity
for the mass function} $M=M_0 v^{n-3}$. These facts show that the third order Lovelock term invites some new
features of the gravitational collapse of a null dust fluid into the game.
Thus, it is worth investigating the effects of higher curvature terms
on the final fate of gravitational collapse.

\acknowledgments{This work has been supported by Research
Institute for Astrophysics and Astronomy of Maragha.
We are very grateful to anonymous referee for
useful comments.}
\begin{center}
\large{\textbf{APPENDIX A}}
\end{center}

In this appendix, we calculate the quasilocal mass in third order
Lovelock gravity. The Misner-Sharp mass  which can be identified
as the quasilocal mass has been introduced in \cite{Mis} for
Einstein gravity and in \cite{Maed1} for Gauss-Bonnet gravity.
The metric of an $n$-dimensional spacetime ($M^{n},g_{\mu \nu }$)
which is a warped product of an ($n-2$)-sphere $S^{n-2}$ and a
two-dimensional orbit spacetime ($M^{2},g_{ab}$) under the
isometries of $S^{n-2}$ may be written as
\[
g_{\mu \nu }dx^{\mu }dx^{\nu }=g_{ab}(y)dy^{b}dy^{b}+r^{2}(y)d\Omega^2,
\]
where $a,b=0,1$; $i,j=2...(n-1)$, and $r$ is a scalar on ($M^{2},g_{ab}$)
with $r=0$ defining its boundary. Using the method of Ref. \cite{Maed2}
for the calculation of the quasilocal mass in Lovelock gravity one obtains:
\begin{eqnarray}
\mathcal{M}&=&\frac{(n-2)V_{n-2}r^{n-7}}{2\kappa^2 _{n}} \Big\{-\Lambda r^{6}%
+r^{4}\left[ 1-(Dr)^{2}\right] +\alpha _{2}r^{2}\left[
1-(Dr)^{2}\right] ^{2}+\frac{\alpha _{3}}{3}\left[
1-(Dr)^{2}\right] ^{3} \Big\} \nonumber\\
&\equiv&\frac{(n-2)V_{n-2}}{2\kappa^2 _{n}}M,  \label{M1}
\end{eqnarray}
where $\alpha _{i}$'s are defined in terms of the Lovelock coefficients $%
\alpha _{i}^{\prime }$ in Sec. (II), $D_{a}$ is a metric compatible linear
connection on ($M^{2},g_{ab}$), $(Dr)^{2}=g^{ab}(D_{a}r)(D_{b}r)$, and $%
V_{n-2}$ is the area of the ($n-2$)-sphere. It is worth mentioning
that the mass (\ref{M1}) is consistent with the quasilocal mass
proposed in Ref. \cite {Maed2}. Also, it is a matter of
straightforward calculations to show that Conjecture 2 of Ref.
\cite{Maed2} is true for third order Lovelock gravity. The
mass given by Eq. (\ref{M1}) for the metric (\ref{metric1})
reduces to
\begin{equation}
M=r^{n-7} \left\{-\Lambda r^{6}%
+r^{4}\left[ 1-f(v,r)\right] +\alpha _{2}r^{2}\left[
1-f(v,r)\right] ^{2}+\frac{\alpha _{3}}{3}\left[ 1-f(v,r)\right]
^{3} \right\}.
\end{equation}
Solving for $f(v,r)$, one obtains the metric function given in Eq.
(\ref{fstat}) which shows that the mass of Eq. (\ref{fstat}) is
equal to the quasilocal mass given by Eq. (\ref{M1}).

\begin{center}
\large{\textbf{APPENDIX B}}
\end{center}

In this appendix, we review the definitions of different types of
trapping horizons, and then consider the problem in third order
Lovelock gravity. To do this, it suitable to write the line
element in the double-null coordinate as
\[
ds^{2}=-2e^{-f}d\zeta ^{+}d\zeta ^{-}+r^{2}d\Omega ^{2},
\]
where $f$ and $r$ are functions of the null coordinates ($\zeta
^{+}$, $\zeta^{-}$). The null vectors $\partial /\partial \zeta
^{\pm }$ will be assumed future-pointing. The expansions may be
defined by
\[
\theta _{\pm}=(n-2)r^{-1}\partial _{\pm }r,
\]
where $\partial _{\pm }$ denotes the coordinate derivative along
$\zeta ^{\pm }$. The expansions measure whether the light rays
normal to a sphere are diverging ( $\theta >0$) or converging
($\theta <0$), or equivalently, whether the area of the spheres is
increasing or decreasing in the null directions \cite{Hay}. Note
that the signs of $\theta _{\pm }$ are geometrical invariants, but
their actual values are not. A compact spatial $(n-2)$-surface is
said to be (i) trapped if $\theta _{+} \theta _{-}>0$, (ii)
untrapped if $\theta _{+} \theta _{-}<0$, and (iii) marginal if
$\theta _{+} \theta _{-}=0$. Without loss of generality, one can
set $\theta _{+} =0$ at a marginal surface, then it is future if
$\theta _{-} <0$, past if $\theta _{-}>0$, bifurcating if $\theta
_{-} =0$, outer if $\partial_- \theta _{+}<0$, inner if
$\partial_- \theta _{+}>0$, and degenerate if $\partial_- \theta
_{+}=0$. The closure of a hypersurface foliated by future or past,
outer or inner marginal surfaces is called a (nondegenerate)
trapping horizon \cite{Hay,Maed3}.

The $\zeta \zeta $-component of the field equation is
\[
\left( \partial _{\pm }^{2}r+\partial _{\pm }f\partial _{\pm }r
\right) \left[ 1+\frac{2\alpha _{2}}{r^{2}}\left( 1+2e^{f}\partial
_{+}r\partial _{-}r\right) +\frac{\alpha _{3}}{r^{4}}\left(
1+2e^{f}\partial _{+}r\partial _{-}r\right) ^{2}\right]
=-\frac{\kappa^2 _{n}}{n-2}T_{\pm \pm},
\]
which reduces to
\[
2\partial _{\pm }\theta _{\pm }\left[ 1+\frac{2\alpha _{2}}{r_{h}^{2}}+\frac{%
\alpha _{3}}{r_{h}^{4}}\right] =-\frac{\kappa^2 _{n}}{n-2}T_{\pm
\pm}
\]
on the trapping horizon. The above discussion shows that, if the
null energy condition holds, the trapping horizon is a future outer
trapping horizon.

Finally, we consider the relation between the null energy
condition and the null convergence condition along the radial null
vectors. It is a matter of calculation to show that for a radial
null vector $k^{\mu} \partial /\partial x^{\mu}= k^{+} \partial
/\partial \zeta^{+}+k^{-} \partial /\partial \zeta^{-}$, one has
\[
\kappa^2 _{n}T_{\mu \nu }k^{\mu }k^{\nu }=R_{\mu \nu }k^{\mu
}k^{\nu }\left\{ 1+\frac{\alpha_2^2}{\alpha_3}\left[
C^{1/3}(r,v)-\gamma ^{1/3}C(r,v)^{-1/3}\right] ^{2}\right\},
\]
for our solution (\ref{fstat}). This equation shows that, as in the case of
Gauss-Bonnet gravity for solutions with a general relativity limit \cite{Maed3},
the null convergence condition
$R_{\mu \nu }k^{\mu }k^{\nu }\geq 0$ is satisfied provided the null energy
condition $T_{\mu \nu }k^{\mu }k^{\nu }>0$ holds.

\end{document}